\documentclass[aps,rmp,reprint,groupedaddress]{revtex4-1}

\usepackage{graphicx}
\usepackage{rotating}
\usepackage{amsmath, amssymb}
\usepackage{bm}
\usepackage{soul,color}

\begin{document}

\title{Efficient inference of protein structural ensembles}
\author{Thomas J. Lane}
\affiliation{Department of Chemistry, Stanford University}
\author{Kyle A. Beauchamp}
\affiliation{Computational Biology Group, Memorial Sloan-Kettering Cancer Center}
\author{Christian R. Schwantes}
\affiliation{Department of Chemistry, Stanford University}
\author{Vijay S. Pande}
\email{Correspondance: pande@stanford.edu}
\affiliation{Departments of Chemistry and Computer Science and Biophysics Program, Stanford University}
\date{\today}

\begin{abstract}
It is becoming clear that traditional, single-structure models of proteins are insufficient for understanding their biological function. Here, we outline one method for inferring, from experiments, not only the most common structure a protein adopts (native state), but the entire ensemble of conformations the system can adopt. Such ensemble models are necessary to understand intrinsically disordered proteins, enzyme catalysis, and signaling. We suggest that the most difficult aspect of generating such a model will be finding a small set of configurations to accurately model structural heterogeneity and present one way to overcome this challenge.
\end{abstract}

\maketitle

\section*{Current Techniques for Elucidating Atomistic Structures are Inherently Limited}

Structures of biomolecules have driven molecular biology and biochemistry ever since Watson and Crick solved the structure of DNA \cite{Crick:1953wm}. Protein and nucleic acid structures provide a mechanistic explanation of biological phenomena, including enzyme catalysis, transfer of genetic information, and molecular signaling. Further, structure plays the medically crucial role of providing a starting point for rational drug design \cite{Banaszak:2000ws}. 

Since the mid-20$^{\mathrm{th}}$ century, however, the methods of obtaining biomolecular structures remain largely unchanged. With the important exception of NMR spectroscopy \cite{Markwick:2008ka}, x-ray crystallography remains the workhorse of structural biology \cite{Banaszak:2000ws}. Because crystals, especially under cryogenic conditions, mostly lack the conformational heterogeneity found at room temperature in the solution phase \cite{Keedy:2014gb, Fraser:2011fy}, the vast majority of high-resolution structures contain little or no information about the structural diversity of proteins.

Such dynamics are implicated in many biological questions currently under investigation, including the role fluctuations (or lack thereof) play in enzyme catalysis \cite{Glowacki:2012jb, Hay:2012eg}, the conformational changes undergone during protein signaling \cite{Bu:2011dy}, and the role of intrinsically disordered proteins in biology \cite{Dyson:2005jb}. The development of new tools to directly study conformational fluctuations could play a key role in answering questions in these fields.

Besides cyro-crystallography, room temperature crystallography and NMR have shown promise as high-resolution methods for the study of structural ensembles \cite{Keedy:2014gb, Fraser:2011fy, Baldwin:2009ds, Korzhnev:2010kj, Sekhar:2012gv}. Both, however, require new methods to rigorously infer ensembles from the experimental readouts obtained. Other techniques such as FRET, PRE, SAXS, IR spectroscopy \cite{Cho:2008fi}, \textit{etc.}~do provide information about the conformational fluctuations of proteins, but are limited in either spatial resolution or total structural information imparted. One could imagine, however, that in the presence of either very strong prior information about the structure of a system, or by combining many of the above techniques into a unified model, enough information might be obtained to infer a high resolution structure. Further, combining structural information from many methods would enable one to profit from the strengths of each, \textit{e.g.}~balancing low resolution global information (as in SAXS) with high-resolution but local probes (FRET).

Many challenges must be overcome to build such integrated structural models. Synthesis of many experimental techniques is not trivial. Further, it can be difficult to rationalize structure in terms of experimental observables, \textit{e.g.}~while NMR chemical shifts contain significant information about a protein sample's structure \cite{Robustelli:2012di}, one cannot simply look at an NMR spectrum and ``see'' even a coarse-grained picture of the structure. Finally, in general, the problem of determining a structure -- much less a structural ensemble -- from available data will be under-determined; there is simply not enough information in even the best experiments to fully specify atomic resolution of an ensemble of structures in a protein system. This makes the deduction of ensemble structural information from the raw data near-hopeless. On the other hand, the inverse problem -- predicting experimental output given a model of the structure -- is generally possible (though sometimes challenging). Therefore, a natural approach to the interpretation of structurally rich experiments is to generate putative models, then test those models for consistency with the data. Many such approaches have been employed in the past \cite{Shen:2008gw, Trabuco:2008cc, Marsh:2009gd, Zhang:2007hf, Choy:2001cr, DeSimone:2009cp, Vendruscolo:2007de}

Here, we outline a way to perform two tasks simultaneously: first, interpret complex experimental data directly as an equilibrium ensemble of structures, and second, rigorously integrate information from many different experiments into a single model. We call this combined project ``integrative modeling''. Previous work along these lines has been pursued and will be discussed momentarily.

\section*{Problem Statement}

Our goal is to extend structural biology by focusing not only on the single most probable equilibrium conformations of macromolecules, but the set of all ``important'' conformations at equilibrium. For such a project to be successful, we need to define specifically what the final model will look like and what difficulties must, in general, be overcome by a method aiming at that endpoint.

The end result of \emph{any} ensemble modeling project will be a set of conformations and their associated Boltzmann weights. The conformations should be important in the sense that they are as representative as possible of the entire Boltzmann distribution, such that for a finite set of structures $\{\mathbf{x}_i\}$, things like expectation values over the Boltzmann distribution are well approximated by a finite sum over those structures. That is, for an arbitrary function of the system coordinates $g(\mathbf{x})$,
\begin{equation}\label{requirement}
\int p(\mathbf{x}) \, g(\mathbf{x}) \, d\mathbf{x} \approx \frac{1}{N(k)} \sum_{i=1}^k p(\mathbf{x}_i) \, g(\mathbf{x}_i)
\end{equation}
Where $N$ is an appropriate normalization constant. We could also envision extending this requirement to higher moments, replacing each $g$ with $g^n$; such a model would capture not only mean properties of the system, but also low-order fluctuations around that mean. We are not suggesting (\ref{requirement}) as a specific metric by which to judge models -- indeed, we never specified $g$ -- but rather as a statement of the kinds of things we would hope ensemble models would get right.

This implies that an ensemble modeling project consists of two parts. First, one needs to be able to assign Boltzmann probabilities to given conformations -- compute $p(\mathbf{x})$ for any $\mathbf{x}$. Second, one needs to generate a discrete set of structures $\{\mathbf{x}_i\}$ such that things like (\ref{requirement}) are well approximated. 

Efficient generation of these structures is the truly hard part of ensemble modeling. One wants to choose a small number of structures in order to keep the model simple and interpretable, but enough structures to satisfy (\ref{requirement}). Clearly, if a relatively small number of structures are to be chosen, the intuition built by (\ref{requirement}) tells us that in the absence of particular constraints, we should choose those that have a high Boltzmann weight $p(\mathbf{x})$. If, however, we know \textit{a priori} that we wish to probe a particular property of the system -- a specific $g(\mathbf{x})$ -- then it might also behoove us to find structures that span the range of values of that function.

Previous methods for ensemble inference have focused on the idea that a set of structures can be generated randomly or from some prior information (such as an empirical forcefield) and then weighted according to some model for $p(\mathbf{x})$, based on experimental data, later on \cite{Fisher:2011bo}. We believe this paradigm inherently limits the efficiency of ensemble inference. If, by chance or good choice, the prior ensemble models the experimental output well, and requires only minimal re-weighting for perfect agreement, then this method is viable. This means, however, our experiments have not substantially furthered our knowledge of the structural heterogeneity of the system. Either the experimental data impart little information, or the original model of the ensemble was already quite good.

Alternatively, the experimental data may be inconsistent with the prior ensemble -- in this case, if the number of structures is small, the ensemble may poorly represent the most stable regions of configuration space. If the number of structures is large, a good model may be obtained, but it is likely that the vast majority of the Boltzmann weight will be concentrated on a small subset of structures. This means that the vast majority of structures are inconsequential, but time was spent computing and evaluating them, an inefficient and unnecessary scheme.

We suggest that, as a default when modeling ensembles, structures should be \emph{chosen} for inclusion in the model based on their Boltzmann weight. This can readily be done by running a Monte Carlo or molecular dynamics simulation in the potential dictated by the model for the Boltzmann weight, $V(\mathbf{x}) \propto - \log p(\mathbf{x})$. There may be cases, such as when one specific system property or observation is particularly important, when this may not be the most efficient scheme. Without additional information, however, it is an excellent place to begin.

Sampling directly from an inferred Boltzmann ensemble opens up the exciting possibility of \emph{directly interpreting experiments} without the need for a specific form of the prior. If structural samples are generated from a prior, then the prior must contain significant information such that it overlaps with ensembles consistent with the experimental information included in the ensemble model. When sampling directly from a posterior Boltzmann distribution, however, this requirement is alleviated. One could imagine running simulations where the prior is a minimal potential, perhaps enforcing only protein primary sequence through bond length and angle constraints. Such a model highlights an important philosophical aspect of the protocol suggested here -- we envision ensemble models as a way to infer structures from experimental data, not just a method for correcting simulation errors.

\subsection*{Previous Work}

While we don't wish to provide a comprehensive review of previous work in both integrating multiple experiments and ensemble modeling, there have been important advances that we build on, and understanding this previous work should help the reader place our ideas in a broader context.\footnote{The authors admit that their coverage of the literature may not be comprehensive, and would very much welcome recommendations for citations to add to this section.}

The foundation for integrative modeling has been well-laid. As early as 2005, Rieping, Habeck and Nilges pointed out that thinking of structure inference in a Bayesian framework could be a fruitful avenue for integrating different experimental techniques into a combined model. This has lead to a body of theoretical and modeling work \cite{Rieping:2005dca, Habeck:2005ft, MacCallum:2014jj}, but the approach has yet to be widely employed to determine structures. Our ideas build directly off theirs, but expands these ideas to modeling entire ensembles rather than single structures, an idea that has lead to a variety of related models \cite{Huang:2010tl, Pitera:2012ds, Fisher:2011bo, Olsson:2013gq,Olsson:2014eg, Beauchamp:2014fy}.

Regarding ensemble modeling, two primary techniques have been thus far pursued. The first is to implement distance restraints into MD simulations to enforce experimental restraints. The second has been to weight or select structures from a pre-selected library of structures to generate an ensemble consistent with experiments. These techniques have successfully been able to reproduce ``test'' ensembles, generated from simulated data (\textit{i.e.}~experimental outcomes are predicted from a simulated ensemble, and then the method attempts to reproduce the ensemble only from these experimental observables). They are nicely reviewed by Fisher and Stultz \cite{Fisher:2011bo}, though work has continued since 2011, \textit{e.g.}~\cite{Olsson:2013gq, Olsson:2014eg}.

Our work demonstrates that these two ideas, both incorporating many experiments in a Bayesian framework and modeling ensembles, can be naturally combined. Further, we suggest that the Bayesian inference necessary to infer an ensemble model can be efficiently conducted by leveraging the well-developed techniques of molecular simulation. Recently, this idea has been implemented to study the structure of simple electrolytes \cite{White:2014fd}.

\section*{Theory}

We gently extend previous work \cite{Rieping:2005dca, Habeck:2005ft, Pitera:2012ds, Beauchamp:2014fy}, to prescribe an efficient protocol for modeling structural ensembles. This work contains two main advances. First, our model accounts for the error in both experiments and prediction algorithms. Second, we suggest that ensemble models be constructed by directly sampling the posterior ensemble, rather than sampling the prior ensemble and then re-weighting.

The result of this procedure is a set of structures and their relative populations that is guaranteed to reproduce the experimental measurements. If the available experimental data are insufficient to uniquely determine an ensemble, the method chooses the ensemble with quantitatively the least amount of information. This guarantees that the calculated ensemble contains only the amount of information contained in the experiments, and no more. Further, this method incorporates experimental uncertainty and error in the prediction of experimental observables through a rigorous Bayesian framework.

The basic idea is as follows. We assume we are given a set of experimental data representing ensemble averaged measurements, along with a set of prediction algorithms that can estimate those experiments given a single structural model. Using these, we can find the \emph{set} of Boltzmann ensembles that reproduce our experiments when we apply those prediction algorithms -- there will not, in general, be a unique ensemble capable of reproducing the experimental data. We choose one amongst these ``degenerate'' ensembles by favoring the one that is closest to some informative prior, which could range from a full-fledged empirical potential (MD forcefield) to a simple model of the bonds of the primary sequence of a protein. This procedure results in a potential energy function that describes the Boltzmann ensemble. Recovering our structural ensemble then consists of nothing more than sampling that potential via some algorithm (\textit{e.g.}~Metropolis Monte Carlo or molecular dynamics) that is guaranteed to reproduce a canonical ensemble.

\subsubsection*{Choosing Amongst Degenerate Ensembles}

Our goal is to produce a structural ensemble that reproduces some experimental observations, taken from a bulk experiment. We ignore momenta and work exclusively in configuration space; a protein configuration is denoted $\mathbf{x}$. Thus, given a set of experimental observations $a_i$, our goal is to produce a Boltzmann distribution $p( \mathbf{x})$ such that those experimental observations are reproduced, \textit{i.e.}~choose $p( \mathbf{x})$ such that $a_i = \langle f_i \rangle$, where
\begin{equation}\label{boltz-avg}
\langle f_i \rangle = \int p( \mathbf{x}) f_i(\mathbf{x}) \> d\mathbf{x}
\end{equation}
and $f_i(\mathbf{x})$ is an algorithm for calculating the experimental observable from a configuration, commonly referred to as a \emph{forward function}.

In general, there may be many such distributions $p( \mathbf{x})$ that reproduce these observables. To choose amongst them we follow Pitera and Chodera \cite{Pitera:2012ds} and choose the ensemble that contains quantitatively the least information possible relative to some prior distribution $m(\mathbf{x})$.\footnote{While the result of our derivation is effectively the same as \cite{Pitera:2012ds} and \cite{Beauchamp:2014fy}, the philosophical choice of minimizing the relative entropy was not highlighted in those works. The derivations there involve an integral involving the differential entropy where the precise connection to the prior potential is obscured.} This additional information can be quantified by the relative entropy, $D$,
\begin{equation}\label{relent}
D \big[ p(\mathbf{x}) || m(\mathbf{x} ) \big] = 
\int p(\mathbf{x}) \log \frac{p(\mathbf{x})}{m(\mathbf{x})} \> d\mathbf{x}
\end{equation}
also called the Kullback-Leibler divergence. Thus, we want the distribution $p(\mathbf{x})$ that minimizes $D$, subject to the constraint that the experimental data are reproduced. In making this choice, we are limiting our ensemble to include \emph{only} information present in the experiments we attempt to match, and no more.\footnote{Thanks to Robert McGibbon for pointing out this is a specific case of posterior regularization \cite{Ganchev:2010ui}.} The relative entropy can be employed \textit{post facto} to quantitatively measure the structural information content of experiments included, and also should provide a mechanism for assessing what additional experiments would be best to perform to continue refining a structural model.

The information gain must be benchmarked against some ``baseline,'' here denoted as the prior distribution $m(\mathbf{x})$ that we choose. A natural choice for $m(\mathbf{x})$ for a situation where large quantities of high-quality experimental information are available might be a potential $V_0(\mathbf{x})$ that includes only bonds, angles, and sterics, enforcing the primary structure of the protein as well as basic geometric constraints (no overlapping atoms), but little more. In the case where the experiments at our disposal are fairly limited, a stronger prior, such as an empirical MD forcefield, could be employed. In principle this potential can be anything from quantum calculations to a non-interacting ideal gas.

We will next demonstrate how to find the distribution $p(\mathbf{x})$ that minimizes (\ref{relent}) subject to the constraint that the Boltzmann distribution obtained gives the experimentally observed values, as in (\ref{boltz-avg}). In the next section, we present a more sophisticated model that includes error models for the experimental results and experimental predictors.

We solve the constrained optimization problem minimizing (\ref{relent}) by introducing the Lagrange functional
\begin{align}\label{lagrangian}
\Lambda \big[ p(\mathbf{x}) \big] =& \int p(\mathbf{x}) \log p(\mathbf{x}) \> d \mathbf{x} \> - \\
&\int p(\mathbf{x}) \log m(\mathbf{x}) \> d\mathbf{x} \> - \notag \\
&\sum_i \lambda_i \left( \int f_i( \mathbf{x} ) p(\mathbf{x}) \> d\mathbf{x} - a_i \right) \notag
\end{align}
where the $\lambda_i$ are Lagrange multipliers. Finding the root of the functional derivative $\delta \Lambda / \delta p$ gives the functional extremum of $p(\mathbf{x})$ (Appendix \ref{appendix_p_lambda}),
\begin{equation}\label{p_lambda}
p_{\bm{\lambda}} (\mathbf{x}) = \frac{1}{Z(\bm{\lambda})} \exp \left\{ \sum_i \lambda_i f_i(\mathbf{x}) - V_0(\mathbf{x}) \right\}
\end{equation}
where  $Z$ is a normalization constant (partition function), we have written the prior in terms of its corresponding potential $m(\mathbf{x}) \propto - \log V_0(\mathbf{x})$, and have added the subscript $\bm{\lambda}$ to emphasize that the specific Boltzmann distribution is parameterized by the Lagrange multipliers. This shows that if we can find the Lagrange multipliers $\bm{\lambda}$, we can sample $p_{\bm{\lambda}} (\mathbf{x})$ by running a simulation in the potential $\sum_i \lambda_i f_i(\mathbf{x}) - V_0(\mathbf{x})$.

The Lagrange multipliers we desire are those that ensure the constraints
\begin{equation}\label{equib-expt}
a_i = \int f_i( \mathbf{x} ) \, p_{\bm{\lambda}}(\mathbf{x}) \> d\mathbf{x}
\end{equation}
are satisfied. This is easily done through an analogy to statistical mechanics \cite{Rieping:2005dca}. Consider the ``free energy'' function
\begin{equation}\label{free-energy-opt}
\Gamma(\bm{\lambda}) = - \log Z(\bm{\lambda}) + \sum_i \lambda_i a_i
\end{equation}
by finding the roots of the derivatives $\partial \Gamma / \partial \lambda_i$ (Appendix \ref{gamma_derivative}), we show that $\langle f_i \rangle_{\bm{\lambda}} = a_i$ for all observables $i$ is one extremum of the function $\Gamma(\bm{\lambda})$. Taking second derivatives shows
\begin{equation}\label{Hessian}
H_{ij} = \frac{\partial^2 \, \Gamma(\bm{\lambda}) }{ \partial \lambda_i \, \partial \lambda_j } = 
\langle f_i \, f_j \rangle_{\bm{\lambda}}
 - \langle f_i \rangle_{\bm{\lambda}} \langle f_j \rangle_{\bm{\lambda}} 
\end{equation}
that the Hessian of $\Gamma(\bm{\lambda})$ is the covariance matrix of the predicted observables $f_i$, which ensures that the Hessian is positive semi-definite. Whenever the the $f_i$ are not linearly dependent, the Hessian will be positive definite. We assume this is the case going forward, and note it can be ensured by intelligently choosing the set of $f_i$ to include in the model.

Since the Hessian of $\Gamma$ is positive definite, this function has a unique extremum, a minimum. At that minimum, $\langle f_i \rangle_{\bm{\lambda}} = a_i$, \textit{i.e.}~the values of $\bm{\lambda}$ satisfy the constraints we wished to find, (\ref{equib-expt}). Thus by numerically minimizing (\ref{free-energy-opt}), a convex problem, we can find the desired Lagrange multipliers $\bm{\lambda}$.

In practice, computing the gradient and Hessian of $\Gamma(\bm{\lambda})$ requires an expectation be evaluated over the equilibrium distribution $p_{\bm{\lambda}}(\mathbf{x})$, which we have written $\langle \cdot \rangle_{\bm{\lambda}}$. This must requires samples of $\mathbf{x}$ to estimate the integral (\ref{equib-expt}). These can be generated by protein Monte Carlo or molecular dynamics, which are guaranteed to produce correct the stationary distribution.

Notice throughout the previous discussion we have not included any treatment of possible error in either the experimental values or in the prediction algorithms $f_i$. Next, we treat such error.

\subsubsection*{Including Error in the Experiment and Prediction Algorithms}

We consider the case where we wish to include a model for the error inherent in both experiment and the forward function predictions in our ensemble model. Suppose we have some model for the probability $P( \mathbf{a} | D )$ that a specific experimental outcome, $\mathbf{a}$, is the \emph{correctly predicted ground truth} given some observed experimental data $D$. The shorthand $P( \mathbf{a} | D )$ includes all error due to \emph{both the experiment and prediction algorithm}. Appendix \ref{paD} provides a general example of how this function might be constructed.

Given a model for $P( \mathbf{a} | D )$, we can treat the possible experimental outcomes as a nuisance parameter, and obtain a single estimate for the Boltzmann distribution
\begin{equation}\label{mixed}
p (\mathbf{x}) = \int_A p_{\bm{\lambda}} (\mathbf{x}) P( \mathbf{a} | D )  \> d \mathbf{a}
\end{equation}
Where $A$ is the set of experiments our forward functions can possibly predict, $A = \{ \mathbf{a} : a_i \in \mathrm{range}[f_i] \}$, excluding \textit{e.g.}~unphysical possibilities such as negative chemical shifts. This model includes \emph{all possible} experimental ``ground truths'', and weights them according to how likely each is given the observed experimental data and the quality of the forward functions.

The integral (\ref{mixed}) cannot be evaluated directly, however, since the Boltzmann distribution for a given experimental ``ground truth'', $p_{\bm{\lambda}}$, is parameterized by $\bm{\lambda}$ and not $\mathbf{a}$. Fortunately, we just derived a method for determining a set of $\bm{\lambda}$ corresponding to a single $\mathbf{a}$ by minimizing the function $\Gamma$. Let us denote this process by $M$, that is $M( \mathbf{a} ) = \bm{\lambda}$. 

Then, we can write a change of variables formula
\begin{equation}\label{change-of-variables}
P( \bm{\lambda} | D) = P( \mathbf{a} | D) \> | J_{M^{-1}} (\bm{\lambda}) | 
\end{equation}
where $J_{M^{-1}}$ is the Jacobian of the inverse of $M$ (Appendix \ref{invertible} proves $M$ is a bijection, and therefore invertible). We can now integrate over $\bm{\lambda}$ instead of $\mathbf{a}$,
\begin{equation}
\label{final-boltz}
p(\mathbf{x}) = \int_L
p_{\bm{\lambda}} (\mathbf{x}) \>
P( \bm{\lambda} | D)  \> | J_{M^{-1}} (\bm{\lambda}) | 
\> d \bm{\lambda} \>
\end{equation}
where $L$ is the image of $A$, $M : A \to L$. In Appendix \ref{Jacobian} we give an explicit formula for $J_{M^{-1}}$.

\section*{Implementation}

We have begun work on an open-source software package named \textsc{Odin} that aims to implement this computational framework in an efficient and accessible manner (\url{www.github.com/tjlane/odin}). 

\textsc{Odin} will orchestrate the following procedure in order to estimate the Boltzmann ensemble, based on the theory above. We wish to efficiently evaluate the integral (\ref{final-boltz}). This can be performed via a Monte Carlo scheme, which will be efficient if we sample regions of $\bm{\lambda}$-$\mathbf{x}$ space where the integrand is large.

To perform this, we suggest the following workflow, which is simple but has no guarantees of optimality:

\begin{itemize}
\item Construct $P(\bm{\lambda} | D)$ and find it's maximum, $\hat{\bm{\lambda}}$, either numerically or analytically.

\item Run extensive protein simulations (MD or MC) in the ensemble $p_{\hat{\bm{\lambda}}}$, and record samples $\{ \mathbf{x}_i \}$. This can be done in parallel by running many independent simulations.

\item Employ Markov chain Monte Carlo to evaluate the high dimensional integral over $\bm{\lambda}$ in $(\ref{final-boltz})$.

\item Run additional sampling or employ left aside samples of $\{ \mathbf{x}_i \}$ to check for convergence of the algorithm. It will be impossible in general to evaluate convergence, but we suggest that sampling can be deemed converged when the ensemble averages of the forward functions employed, $\langle f_i \rangle$, stabilize.

\end{itemize}

One possibility that may be fruitful to explore would be running many different simulations in different $\bm{\lambda}$ ensembles, perhaps clustered around $p_{\hat{\bm{\lambda}}}$, or by changing the values of $\bm{\lambda}$ in serial as simulations progress. Such schemes are directly analogous to the various replica-based and annealing approaches employed in molecular simulation, but here the thermodynamic variables are the Lagrange multipliers $\bm{\lambda}$ (instead of \textit{e.g.}~temperature). This could speed sampling, but which scheme will be most efficient is very difficult to analyze and will require heuristics derived from experience.

\subsection*{A Comment on Efficiency}

A major shortcoming of our scheme is that the factors $Z(\bm{\lambda})$ and $J_{M^{-1}} (\bm{\lambda})$ require integrals over the space of all $\mathbf{x}$ in order to evaluate. This greatly reduces the potential efficiency of the method, as samples of $\mathbf{x}$ are necessary to evaluate and sample $p(\mathbf{x})$, leading to the need for an iterative scheme. The workflow suggested in the previous section calls for a single iteration, the minimum required, but it is currently unknown if this is the most efficient protocol possible.

Sadly, this problem is general to all inference schemes that attempt to model ensemble properties (such as expectations). Consider properties such as (\ref{boltz-avg}). If we wish to ensure that the mean value of an observable, $\langle f_i \rangle$, has a specific value without any \textit{a priori} knowledge of the structure of $f_i$ or the value of the desired expectation $\langle f_i \rangle$, then \emph{it will be required that we sample values of $f_i(\mathbf{x})$ in order to evaluate the viability of the Boltzmann distribution $p(\mathbf{x})$}. This problem will be faced by any scheme that aims to reproduce ensemble properties.

We suspect that the most efficient ensemble modeling algorithms will be those equipped with efficient ways to sample both $\mathbf{x}$ and any auxiliary parameters (here, $\bm{\lambda}$) simultaneously in a way guaranteed to concentrate on regions of configuration space where the Boltzmann weight $p(\mathbf{x})$ is large. Precisely how to do this for the present method is a subject of ongoing research.

\section*{Future Directions}

We have presented a method that enables the inference of ensemble structural models from diverse experiments. Our method has a number of attractive properties, including:
\begin{itemize}
\item the ability to model arbitrary experimental observables,
\item a rigorous treatment of experimental and forward function error, and
\item the capability to sample directly from a posterior Boltzmann probability
\end{itemize}
The method presented here is not unique in this regard, but we feel these three properties form a minimum set of requirements for ensemble modeling techniques, and thus believe the method presented here a viable option for ensemble modeling.

A number of further avenues remain for further research in the nascent field of ensemble modeling. Any such model will be limited by the quality of the forward functions employed. Research into efficient, accurate predictors is mandatory. Software implementations of these methods and standardized ways of processing experimental data need to be developed and curated by a nascent community.

We have suggested that methods employing long MC or MD runs will be necessary to compute expectations of the form (\ref{equib-expt}), which might make such calculations intractable for a large portion of the structural biology community. The good news is that by leveraging developments from protein simulation community, advances in hardware, software and equilibrium sampling techniques can be leveraged to speed this process; it should be tractable at present for small systems (100 residues or less).

Finally, as discussed previously, research remains to be done into efficient methods for sampling from $p(\mathbf{x})$ directly. Such sampling schemes will alleviate the need for highly informative priors and should generate large gains in modeling efficiency. We hope that methods for performing MCMC sampling, developed in statistics, might be reappropriated for this purpose.

These challenges granted, we believe this scheme or some other closely related one holds great hope for the integrative modeling of structural ensembles.

\begin{acknowledgments}
We are grateful to John Chodera for sharing his notes on this subject matter, which led quickly to the extension of his work with Jed Pitera \cite{Pitera:2012ds} to include error. Discussions with Seb Doniach, Rhiju Das, and Robert McGibbon have also been invaluable.
\end{acknowledgments}

\appendix 

\section{Details of the Derivation of $p_{\bm{\lambda}}$}

\subsection{Derivation of Equation (\ref{p_lambda})}
\label{appendix_p_lambda}

We begin with the Lagrange functional (\ref{lagrangian}). We wish to find the functional extremum. Analogous to standard (non-functional) Lagrange multipliers, taking derivatives with respect to the conjugate variables $\lambda_i$, $0 = \partial \Lambda / \partial \lambda_i$, reproduces the constraints (\ref{boltz-avg}). The functional derivative $\partial \Lambda / \partial p$ is defined as
\[
\int \frac{\partial \Lambda [p] }{\partial p ( \mathbf{x} )} \> \phi ( \mathbf{x} ) \> d \mathbf{x} =
\frac{d}{d \epsilon} \Lambda [ p + \epsilon \phi ] \Big|_{\epsilon=0}
\]
evaluating the left hand side,
\begin{align*}
\frac{d}{d \epsilon} \Lambda [ p + \epsilon \phi ] \Big|_{\epsilon=0} =&
\int \phi + \phi \log p \> d\mathbf{x} \\
&- \int \phi \log m \> d\mathbf{x} \\
&- \sum_i \lambda_i \int f_i \, \phi \> d\mathbf{x}
\end{align*}
which, by the functional derivative definition, shows
\[
\frac{\partial \Lambda [p] }{\partial p} = 1 + \log p - \log m - \sum_i \lambda_i f_i
\]
at the extremum $\partial \Lambda / \partial p = 0$,
\[
\log p = \log m + \sum_i \lambda_i f_i -1
\]
Because the resulting $p$ is parameterized in terms of $\bm{\lambda}$, and is not normalized, we will denote its normalized form by $p_{\bm{\lambda}} \equiv p / \int p \> d\mathbf{x}$. This allows us to write
\begin{equation}
p_{\bm{\lambda}} (\mathbf{x}) \propto m(\mathbf{x}) \exp \left\{ \sum_i \lambda_i f_i(\mathbf{x}) - 1\right\}
\end{equation}
Defining a normalization constant (partition function) $Z(\bm{\lambda}) = e^{-1} \int m(\mathbf{x}) \exp \left\{\sum_i \lambda_i f_i(\mathbf{x})  \right\} \> d\mathbf{x}$, we obtain
\begin{equation}
p_{\bm{\lambda}} (\mathbf{x}) = \frac{m(\mathbf{x})}{Z(\bm{\lambda})} \exp \left\{ \sum_i \lambda_i f_i(\mathbf{x}) \right\}
\end{equation}
Finally, rewriting $m(\mathbf{x}) = - \log V_0 (\mathbf{x})$ gives (\ref{p_lambda}).

\subsection{Derivation of $\nabla \Gamma$}
\label{gamma_derivative}

Beginning with (\ref{free-energy-opt}), we show $\langle f_i \rangle_{\bm{\lambda}} = a_i$ is a local extremum of $\Gamma$. A similar calculation can be used to derive the Hessian $H(\bm{\lambda})$ and later the Jacobian $J_{M^{-1}}$ by taking second derivatives.
\begin{align}
0 &= \frac{\partial \Gamma}{\partial \lambda_i} \notag \\
&= - \frac{1}{Z(\bm{\lambda})} \frac{\partial Z(\bm{\lambda})}{\partial \lambda_i} + a_i \ \notag \\
&= - \frac{1}{Z(\bm{\lambda})}  \frac{\partial}{\partial \lambda_i} \left( \int \exp \left\{  \sum_i \lambda_i f_i(\mathbf{x}) - V_0(\mathbf{x}) \right\} d \mathbf{x} \right) + a_i \notag \\
&= - \frac{1}{Z(\bm{\lambda})} \int f_i(\mathbf{x}) \exp \left\{  \sum_i \lambda_i f_i(\mathbf{x}) - V_0(\mathbf{x}) \right\} d \mathbf{x} + a_i \notag \\
&= - \langle f_i \rangle_{\bm{\lambda}} + a_i \notag
\end{align} 
This calculation may be familiar from statistical mechanics, as it is the method employed to find equilibrium values of thermodynamic parameters by minimizing a free energy function.

\subsection{Explicit form of the Jacobian}
\label{Jacobian}

We compute the Jacobian $J_{M^{-1}}$ for reference. One element of the Jacobian matrix can be written
\begin{align*}
\left[ J_{M^{-1}} \right]_{ij} &= \frac{\partial M^{-1}_i (\bm{\lambda})}{\partial \lambda_j} \\
&= \frac{\partial}{\partial \lambda_j}
\left( \int \frac{f_i(\mathbf{x})}{Z(\bm{\lambda})} \exp \left\{  \sum_i \lambda_i f_i(\mathbf{x}) - V_0(\mathbf{x}) \right\} d \mathbf{x} \right) \\
&= \langle f_i \, f_j \rangle_{\bm{\lambda}}
 - \langle f_i \rangle_{\bm{\lambda}} \langle f_j \rangle_{\bm{\lambda}} 
\end{align*}
Note this is the exact same computation as in (\ref{Hessian}).

\subsection{Proof M is Invertible}
\label{invertible}

We demonstrate that the process of translating a set of experimental observations $\mathbf{a}$ into a set of Lagrange multipliers $\bm{\lambda}$ is bijective for possible values of $\mathbf{a}$. 

We denote this procedure by $M( \mathbf{a} ) = \bm{\lambda}$; previously, we showed that this could be done numerically by minimizing $\Gamma$ in eq.~(\ref{free-energy-opt}). Recall $A = \{ \mathbf{a} : a_i \in \mathrm{range}[f_i] \}$, and $L$ is the image of $A$ under operation $M$. Then we will show $M: A \to L$ is bijective so long as the Hessian (\ref{Hessian}) is positive definite everywhere in $\bm{\lambda}$.

For a given $\bm{\lambda} \in L$, we know the corresponding $\mathbf{a}$ is given by $\langle \mathbf{f} \rangle_{\bm{\lambda}}$. Suppose there existed $\mathbf{a}'$ such that $M(\mathbf{a}') = M(\mathbf{a})$. Since the Hessian of $\Gamma$ in  (\ref{Hessian}) is positive definite, then $M(\mathbf{a}') = M(\mathbf{a}) = \bm{\lambda}$. Further, it would also be the case that $\langle \mathbf{f} \rangle_{\bm{\lambda}} = \mathbf{a}' = \mathbf{a}$. This shows that $M$ is one-to-one. Since $L$ is defined as the range of $M$, it is also onto. Therefore, $M$ is a bijection, and has an inverse.

\section{Example of Constructing $P(\mathbf{a} | D)$}
\label{paD}

We stated previously that the error model written $P(\mathbf{a} | D)$ should include errors for both the experiment and prediction models employed. It is generally straightforward to write down the experimental error model as the probability of a specific ground truth $\mathbf{a}'$ given some experimental data, $D$. Prediction algorithms also usually report the probability that a given value $\mathbf{a}$ is the ground truth given the prediction was some other value, say $\mathbf{a}'$.

These two error models can then be composed into a single model
\[
p(\mathbf{a} | D) = \int P_f(\mathbf{a} | \mathbf{a}') \, P_e(\mathbf{a}' | D) \> d\mathbf{a}'
\]
where $P_f(\mathbf{a} | \mathbf{a}')$ and $P_e(\mathbf{a}' | D)$ represent the error models for the predictors and experiments, respectively.

Let us provide a concrete example of this construction. First we consider $P_e(\mathbf{a}' | D)$. Suppose the experiments are well-modeled by a multivariate normal. Then we could write a likelihood function
\[
\mathcal{L} (D | \bm{\mu}, \Sigma) \propto \prod_{\mathbf{d} \in D}
\exp \left\{ - \frac{1}{2} (\mathbf{d} - \bm{\mu})^T \Sigma^{-1} (\mathbf{d} - \bm{\mu}) \right\}
\]
If we assume the parameters $\bm{\mu}$ and $\Sigma$ have uniform priors, then we can invoke Bayes' theorem to write
\[
P_e(\mathbf{a}' | D) \propto  \iint \mathcal{L} ( \mathbf{a}' | \bm{\mu}, \Sigma ) \mathcal{L} (D | \bm{\mu}, \Sigma) \> d\bm{\mu} \> d\Sigma
\]

Further suppose that we have a model of the error of the predictors $\{ f_i \}$, $P_f(\mathbf{a} | \mathbf{a}')$. To continue our concrete example, consider the basic case where the experimental predictors have a Gaussian spread, defined by covariance matrix $S$, whatever the predicted value may be. Then we could write
\[
P_f (\mathbf{a} | \mathbf{a}') \propto
\exp \left\{ - \frac{1}{2} (\mathbf{a} - \mathbf{a}')^T S^{-1} (\mathbf{a} - \mathbf{a}') \right\}
\]

By combining the previous steps, we could construct our final model for $P( \mathbf{a} | D )$, up to a normalization constant $Z$,
\[
P( \mathbf{a} | D ) = \frac{1}{Z} \int P_f(\mathbf{a} | \mathbf{a}') \mathcal{L}(\mathbf{a}' | \bm{\mu}, \Sigma)
\mathcal{L}(D | \bm{\mu}, \Sigma)
\> d\bm{\mu} \> d\Sigma \> d\mathbf{a}'
\]
such a model could be evaluated numerically \textit{e.g.}~by MCMC.

This provides an illustration the construction of $P( \mathbf{a} | D )$; the specifics of this model will necessarily vary based on the application at hand.

%----------------------------------------------------------------------------------------
%	REFERENCE LIST
%----------------------------------------------------------------------------------------

%\newpage
\bibliography{papers2.bib}

%----------------------------------------------------------------------------------------

\end{document}